\documentclass[preprint,superscriptaddress,prd,showpacs]{revtex4}
\usepackage{amssymb,amsmath,graphicx,amscd,xcolor}

\begin{document}

\title{Quantum Electric Field Fluctuations and Potential Scattering}

\author{Haiyun Huang}
\email{haiyun.huang@tufts.edu}
\affiliation{Institute of Cosmology, Department of Physics and Astronomy \\
Tufts University, Medford, Massachusetts 02155, USA}

\author{L. H. Ford}
\email{ford@cosmos.phy.tufts.edu}
\affiliation{Institute of Cosmology, Department of Physics and Astronomy \\
Tufts University, Medford, Massachusetts 02155, USA}

\begin{abstract}
Some physical effects of time averaged quantum electric field fluctuations are discussed. 
The one loop radiative correction to potential scattering are approximately derived from  
simple arguments which invoke vacuum electric field fluctuations. For
both above barrier scattering and quantum tunneling, this effect increases the transmission
probability. It is argued that the shape of the potential determines a sampling function for the
time averaging of the quantum electric field operator. We also suggest that there is a
nonperturbative enhancement of the transmission probability which can be inferred from the
probability distribution for  time averaged electric field fluctuations.
 \end{abstract}
 

\maketitle
\baselineskip=14pt	

\section{Introduction}
\label{sec:intro}

The vacuum fluctuations of the quantized electromagnetic field give rise a number of physical
effects, including the Casimir effect, the Lamb shift, and the anomalous magnetic moment of the
electron. Many of these effects are calculated in perturbative quantum electrodynamics, often
by a procedure which does not easily lend itself to an interpretation in terms of field fluctuations.
An exception is Welton's~\cite{Welton} calculation of the dominant contribution to the Lamb
shift, which leads to a simple physical picture in which electric field fluctuations cause an electron
in the $2s$ state of hydrogen to be shifted upwards in energy. One of the purposes of this paper
will be to seek additional examples of this type.

It is well known that time averaging of quantum fields is needed to produce mathematically well 
defined operators. Usually, a test function of compact support is employed for this purpose~\cite{PCT}.
However, in rigorous quantum field theory, this is a formal device which is not given a physical
interpretation. In a recent paper~\cite{BDFS14}, it was suggested that the test, or sampling functions
can have a physical meaning. This paper deals with the propagation of pulses in nonlinear
optical materials and suggests that time averaged vacuum fluctuations of the electric field can alter 
the pulse propagation time inside the material. Furthermore, Ref.~\cite{BDFS14} hypothesizes that
the sampling function is determined by the geometry of the nonlinear material. In the present paper,
we will explore this hypothesis in a different context, that of electron scattering by a potential barrier.

We will be concerned with the vacuum fluctuations of the electric field in a particular direction. Let
${\cal E}({\bf x},t)$ be a Cartesian component of the quantum electric field, such as the $x$-component. We
wish to average this operator over a timelike world line. By going to the rest frame of an observer
moving on this worldline, the averaging can be in time alone at a fixed spatial coordinate. Let
$f_\tau(t)$ be a sampling function of characteristic width  $\tau$, whose time integral is unity
\begin{equation}
\int_{-\infty}^\infty f_\tau(t) \, dt =1 \,.
\label{eq:fnorm}
\end{equation} 
We define the averaged electric field component by
\begin{equation}
\bar{\cal E} = \int_{-\infty}^\infty  {\cal E}({\bf x},t) \, f_\tau(t) \, dt \,.
\label{eq:t-ave}
\end{equation} 
Both ${\cal E}({\bf x},t)$ and $\bar{\cal E}$ have a vanishing mean value in the vacuum state
\begin{equation}
\langle 0| \bar{\cal E} |0 \rangle = 0 \,.
\label{eq:E1}
\end{equation}  
However, $\langle 0| {\cal E}^2({\bf x},t) |0 \rangle$ is infinite, while mean squared value of the averaged field
is finite
\begin{equation}
\langle 0| \bar{\cal E}^2 |0 \rangle = \frac{\eta^2}{\tau^4} \,,
\label{eq:E2}
\end{equation}
where $\eta$ is a dimensionless constant determined by the explicit form of  the sampling function.
(Lorentz-Heaviside units with $c=\hbar =1$ will be used here, so the  electric field has dimensions of 
inverse time squared or inverse length squared.) Note that the mean squared value of the averaged
electric field scales as $1/\tau^4$, so shorter sampling times lead to larger fluctuations due to the 
contribution of higher frequency modes. Typical values of $\eta$ are somewhat less than one. For
example, $\eta = 1/(\sqrt{3} \pi)$ for the Gaussian sampling function,  
$f_\tau(t) = \exp(-t^2/\tau^2)/(\sqrt{\pi}\, \tau)$.

It is convenient to define a dimensionless variable by
\begin{equation}
\chi =  \bar{\cal E}\, \tau^2\,.
\end{equation}
The moments of $\chi$, and hence of $\bar{\cal E}$, are those of a Gaussian probability distribution:
\begin{equation}
P(\chi)=\frac{1}{\sqrt{2\pi}\, \eta}\,\exp\left(-\frac{\chi^2}{2\eta^2}\right) \,.
\label{eq:pdf}
\end{equation}
Now Eq.~(\ref{eq:E2}) gives the second moment of this distribution
\begin{equation}
\langle \chi^2 \rangle = \int_{-\infty}^{\infty} \chi^2 \, P(\chi)\, d\chi = \eta^2\,.
\end{equation}
 Equation~(\ref{eq:pdf}) is the familiar result that the fluctuations of a free quantum field are Gaussian.
 
 As for other quantum field fluctuations, vacuum electric field fluctuations are strongly anticorrelated.
 This means that a fluctuation on a time scale $\tau$ is likely to be followed by a fluctuation of the
 opposite sign. This anticorrelation prevents Brownian motion of a charged particle in the vacuum~\cite{YF04},
 as is required by energy conservation. However, a time dependent background can upset the anticorrelations, 
 and allow the particle to gain average energy~\cite{BF09,PF11}. The key point is that energy conservation is 
 only required on longer 
 time scales, and quantum fluctuations can temporarily violate energy conservation on scales consistent with
 the energy-time uncertainty principle. Here we will be examining a situation where these temporary
 violations of energy conservation can lead to observable effects.
 
 The outline of this paper is as follows: Section~\ref{sec:QM} will review a perturbative treatment of quantum
 scattering in one space dimension, where the incident particle energy exceeds the height of the potential
 barrier. The one loop QED correction to this scattering will be summarized in Sec.~\ref{sec:rad}. In 
 Sec.~\ref{sec:eflucts}, we will present an order of magnitude rederivation of this one loop correction
 based upon vacuum electric field fluctuations, and argue that this provides a simple physical picture for
 the origin of the QED correction.  In Sect.~\ref{sec:tunnel}, we repeat this discussion of one loop
 corrections for the case of quantum tunneling. In  Sect.~\ref{sec:nonpert}, we propose a nonperturbative 
 correction to the tunneling rate arising from large but rare electric field fluctuations. Our results are
 summarized and discussed in Sec.~\ref{sec:sum}.

\section{Radiative Corrections to Above Barrier Potential Scattering}
\label{sec:above}

\subsection{Quantum Scattering in One Space Dimension}
\label{sec:QM}

\begin{figure}
 \centering
 \includegraphics[scale=0.8]{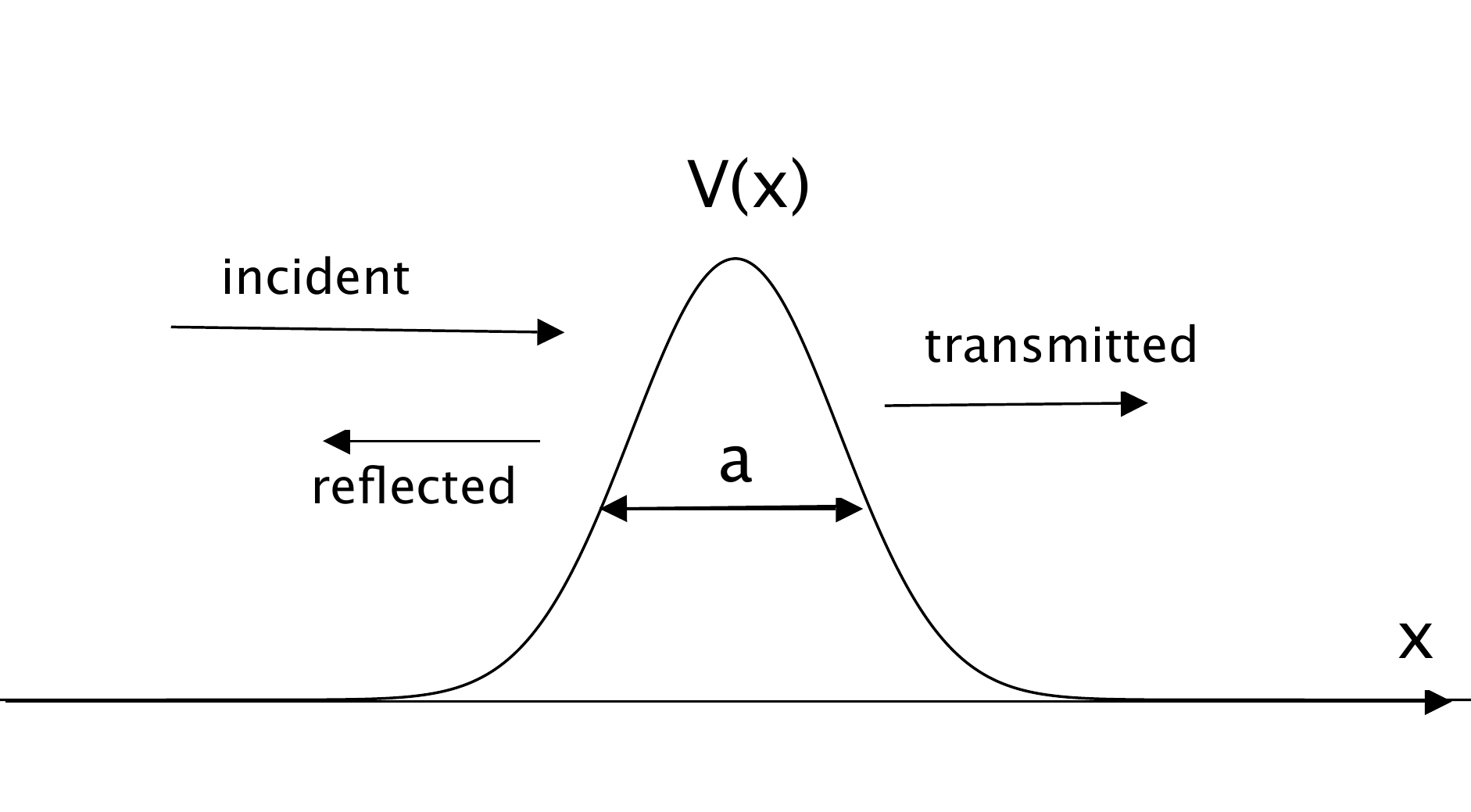}
 \caption{Scattering in one space dimension by a potential $V(x)$ with characteristic width $a$
 is illustrated. A particle is incident from the left, and may either be transmitted or reflected back to the left. }
 \label{fig:V}
 \end{figure}

Consider the scattering of a nonrelativistic electron by a potential $V(x)$ in one space dimension, which is
illustrated in Fig.~\ref{fig:V}. Here we assume that the incident energy of the electron $E_0$ is large
 compared to the maximum of the potential, so the scattering may be treated perturbatively.
If $E_0 = p_0^2/(2 m)$, we can write the one dimensional time independent 
Schr{\"o}dinger equation as
\begin{equation}
\psi''(x) +p_0^2 \psi(x) = 2m \, V(x)\, \psi(x)\,.
\end{equation}
Here $p_0 = m v_0$ is the incident momentum, $v_0$ is the speed, and $m$ is the mass.
This equation is equivalent to the integral equation
\begin{equation}
\psi(x) = \psi_0(x) + \int_{-\infty}^{\infty} G(x-x')\, V(x')\, \psi(x')\, dx'\,,
\label{eq:inteq}
\end{equation}
where $\psi_0(x)$ is a solution of the free Schr{\"o}dinger equation, with $V(x) =0$, and
$G(x-x')$ is a Green's function which satisfies
\begin{equation}
\frac{d^2 G(x-x')}{dx^2} + p_0^2\, G(x-x') = 2 m \, \delta(x-x')\,.
\end{equation}
The explicit form of this Green's function can be taken to be
\begin{equation}
   G(x-x')=\left\{
   \begin{array}{rl}
      -i\frac{m}{p_0}e^{+ ip_0(x-x')}, &x>x'\\
      -i\frac{m}{p_0}e^{- ip_0(x-x')}, &x<x'
   \end{array}
   \right.
\end{equation}
The perturbative solution for $\psi$ to a given order is obtained by iteration of Eq.~(\ref{eq:inteq}).

Here we need only the first order solution, obtained by replacing $\psi$ by  $\psi_0$ in
the right hand side of Eq.~(\ref{eq:inteq}). Let $\psi_0(x) = {\rm e}^{i p_0 x}$, corresponding to a
particle incident on the barrier from the left. The first order solution for a particle  reflected
back to the left is
\begin{equation}
\psi_1(x) = r\,   {\rm e}^{-i p_0 x} \,,
\end{equation}
where $r$ is the reflection amplitude given by
\begin{equation}
 r=-i\frac{m}{p_0} \, \int_{-\infty}^{\infty}   dx\, V(x)\,  {\rm e}^{2 i p_0 x}\,.
 \label{eq:r}
\end{equation}
The reflection and transmission probabilities are given by $R = |r|^2$ and $T=1-R$, respectively.
Note that the transition matrix element is
\begin{equation}
M_0 = \langle -p_0| V |p_0 \rangle = \int_{-\infty}^{\infty}   dx\, V(x)\,  {\rm e}^{2 i p_0 x}\,,
\label{eq:M0}
\end{equation}
and the factor proportional to $m/p_0 = 1/v_0$ in $r$ is a kinematic factor. The reflection probability
becomes
\begin{equation}
R = \frac{|M_0|^2}{v_0^2}\,.
\label{eq:R}
\end{equation}

\subsection{Radiative Corrections to Scattering}
\label{sec:rad}

\begin{figure}
 \centering
 \includegraphics[scale=2.0]{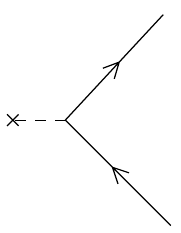}
 \caption{The tree level diagram which describes scattering by an external potential at the level
 of quantum mechanics.}
 \label{fig:tree}
 \end{figure}

\begin{figure}
 \centering
 \includegraphics[scale=2.0]{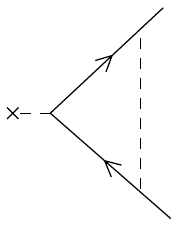}
 \caption{The one loop vertex correction to potential scattering.}
 \label{fig:vertex}
 \end{figure}

Here we discuss the one loop quantum electrodynamic corrections to potential scattering. In Feyman
diagrams, the lowest order nonrelativistic scattering reviewed in the previous subsection corresponds
to Fig.~\ref{fig:tree}. The one loop correction to this process of interest here is the vertex correction,
described by the diagram in Fig.~\ref{fig:vertex}.
The computation of the vertex correction is discussed in many references, such as Ref.~\cite{JR}.
The result is that the transition matrix element, $M_0$ is modified to  $M_0 + M_V$, where
 \begin{equation}
M_V = -\frac{\alpha}{3\pi}\,\frac{q^2}{m^2}\left[\ln\left(\frac{m}{\lambda}\right)+\text{const of order 1}\right]\, M_0\,,
\label{eq:MV}
\end{equation}
where $\alpha= e^2/(4\pi)$ is the fine structure constant,  $q$ is the four-momentum transfer in the scattering
process, and $\lambda$ is an infrared cutoff. A spacelike metric was assumed in writing Eq.~(\ref{eq:MV}). 
Here we are concerned with elastic scattering, so $q$ is a spacelike vector with zero time component in
the rest frame of the potential, and $q^2 = 4 p_0^2$.

\begin{figure}
 \centering
 \includegraphics[scale=2]{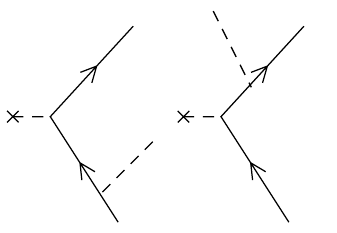}
 \caption{Bremsstrahlung processes in which soft photons are emitted during scattering.}
 \label{fig:brem}
 \end{figure}

The infrared cutoff $\lambda$ arises because the vertex diagram, Fig.~\ref{fig:vertex}, contains an infrared
divergence. The solution to this problem is well known, and first given by Bloch and Nordsieck~\cite{BN}.
It involves the realization that in any scattering process, there is a nonzero probability of emission of soft
photons by bremsstrahlung, illustrated in Fig.~\ref{fig:brem}. In any experiment, there is a threshold of
photon energy below which the soft photons cannot be detected, so the bremsstrahlung process becomes
indistinguishable from the scattering process. However, this is a practical limitation rather than a fundamental
one, so the  bremsstrahlung and scattering processes should be added incoherently. The squared matrix
element for the bremsstrahlung process can be written as
\begin{equation}
|M_b|^2 = \frac{2\alpha}{3\pi}\frac{q^2}{m^2}\left[\ln\left(\frac{2\Delta \epsilon}{\lambda}\right)-\text{const of order 1}\right]
\, |M_0|^2\,,
\end{equation}
where $\bigtriangleup\epsilon$ is the lowest energy we can detect for the soft photon. The net squared
matrix element for scattering, including bremsstrahlung, becomes
 \begin{equation}
|M_0 + M_V|^2 + |M_b|^2 =  |M_0|^2\, 
\left\{1-\frac{2\alpha}{3\pi}\frac{q^2}{m^2}\left[ \ln\left(\frac{m}{2\Delta \epsilon}\right)+\text{const of order 1}\right]\right\}\,.
\end{equation}
Note that the infrared cutoff, $\lambda$, no longer appears.  

The net effect of the radiative correction to the quantum scattering by the potential $V(x)$ is to decrease the reflection
probability and hence increase the transmission probability. The decrease in reflection probability can be
written as
\begin{equation}
\delta R = -\frac{2 e^2\, v_0^2}{3 \pi^2} \, \ln\left(\frac{m}{2\Delta \epsilon}\right) \; R \,,
\end{equation}
where $v_0 =p_0/m$ is the speed of the electron before and after scattering. Note that the magnitude of this
decrease grows quadratically with increasing electron speed. Although the logarithmic factor is formally
divergent as $\Delta \epsilon \rightarrow 0$, it grows very slowly and can never be very large in
a realistic experiment. For example, in an experiment performed at finite temperature $T$, we need 
$\Delta \epsilon \geq k_B T$, where $k_B$ is Boltzmann's constant, in order that the bremsstrahlung
photons can be distinguished from the thermal photons. If $m$ is the electron mass, and $T = 300K$, then
\begin{equation}
 \frac{2}{3 \pi^2}  \, \ln\left(\frac{m}{2\Delta \epsilon}\right) \approx 1.09\,, 
\end{equation}
and is only slightly larger ($1.38$) at $T = 4K$. Here we set $\Delta \epsilon = k_B T$ in
both cases. This implies that a reasonable approximation is
\begin{equation}
\delta R \approx - e^2\, v_0^2\, R\,.
\label{eq:delR}
\end{equation}

\subsection{Electric Field Fluctuations}
\label{sec:eflucts}

Here we wish to give an alternative derivation of Eq.~(\ref{eq:delR}) using a simple physical picture invoking
vacuum fluctuations of the electric field. If $E_0 \gg V(x)$, then the speed of the electron as it passes over the
barrier is approximately its initial speed, $v_0$. If the characteristic width of the barrier is $a$, then the time
required to transit over the barrier is  $\tau \approx a/v_0$. Now we assume that the electron effectively samples 
the quantized electric field with a sampling function $f_\tau(t) \propto V(v_0 t)$. This implies that the electron
feels a mean electric field of about $\bar{\cal{E}} =\eta/\tau^2$, and a force of order $e \bar{\cal{E}}$, which
changes the electron's momentum during the transit by
\begin{equation}
\delta p = e \bar{\cal{E}} \, \tau = \frac{e \eta}{\tau} = \frac{e \eta v_0}{a}\,.
\label{eq:del-p}
\end{equation}
Although the electrons are quantum particles in wavepacket states, we assume that Ehrenfest's theorem 
allows us to use Newtonian mechanics to calculate $\delta p$, which will only be used to find averages
or averages of a square, but not the properties of an individual electron. The sign of the change in
Eq.~(\ref{eq:del-p}) can be either positive or negative with equal probability. Thus we need to examine
in more detail how a small change in momentum during transit over the barrier changes the reflection
probability. Recall that quantum field fluctuations are strongly anticorrelated, so the change in momentum
during the transit is likely to be quickly reversed by a fluctuation in the opposite direction soon after the 
electron has cleared the barrier. Our key assumption will be that the kinematic factor proportional to
$m/p_0$ in Eq.~(\ref{eq:r}) is not sensitive to the temporary change in momentum, but the matrix element $M_0$
is sensitive to it. Under this assumption, we may use Eq.~(\ref{eq:R}) to compute the change in reflection 
probability as being proportional to the change in the squared matrix element, $|M_0|^2$,
\begin{equation}
\delta R = \frac{m^2}{p_0^2}\, \delta |M_0|^2 \,.
\label{eq:delR-2}
\end{equation}
We may find $\delta |M_0|^2$ by a Taylor expansion near $p=p_0$,
\begin{equation}
\delta |M_0|^2 = \frac{\partial (|M_0|^2)}{\partial p_0} \delta p + 
\frac{1}{2} \frac{\partial^2 (|M_0|^2)}{\partial p_0^2} (\delta p)^2 + \cdots \,.
\end{equation}
If we average on $\delta p$, the linear term in the above expression will vanish. 

We need to compute the second derivative of the squared matrix element, using  Eq.~(\ref{eq:M0}).
We may simplify the discussion by assuming that the potential $V(x)$ is even and that its
width is small compared to $1/p_0$, so that 
\begin{equation}
V(-x) =V(x)\,, \quad a \ll \frac{1}{p_0} \,.
\end{equation}
Now we have
\begin{equation}
M_0(p_0) = \int_{-\infty}^{\infty}   dx\, V(x)\,  \cos(2  p_0 x) 
\approx \int_{-\infty}^{\infty}   dx\, V(x)\, (1- 2 p_0^2\, x^2)\,.
\end{equation}
Let
\begin{equation}
\int_{-\infty}^{\infty}   dx\, x^2\,V(x) = \xi a^2\, \int_{-\infty}^{\infty}   dx\, V(x)\,,
\end{equation}
where we expect $\xi$ to be a positive constant slightly less than one.
To leading order, we have
\begin{equation}
M_0' = -4 \xi p_0 a^2\, M_0\,, \qquad M_0'' = -4 \xi a^2\, M_0 \,,
\end{equation}
where here prime denote a derivative with respect to $p_0$. Thus
\begin{equation}
(M_0^2)'' = 2( M_0'' M_0 + M_0' M_0') \approx -8 \xi a^2\, M_0^2 \,.
\label{eq:dp}
\end{equation}

It is useful to examine a few specific choices for $V(x)$ here. For a square barrier,
\begin{equation}
   V(x)=\left\{
   \begin{array}{rl}
      V_0, &|x| \leq a/2\\
       0, & |x| > a/2
   \end{array}
   \right.
\end{equation}
we find from Eq.~(\ref{eq:M0})
\begin{equation}
M_0 = V_0 \frac{\sin(p_0a)}{p_0}\,.
\end{equation}
When $p_0 a \ll 1$, this leads to Eq.~(\ref{eq:dp}) with $\xi = 1/12$.
Note that the square barrier does not lead to 
a suitable choice of sampling function $f_\tau(t)$ because it is not sufficiently
differentiable, but for the purpose of calculating scattering amplitudes, it is a
good approximation to a smooth, flat topped potential. Another example is a
Gaussian form for the potential,
\begin{equation}
V(x)= V_0 \,  {\rm e}^{-(2 x/a)^2}\,,
\end{equation}
leading to Eq.~(\ref{eq:dp}) with $\xi = 1/8$.

Now we may complete our heuristic derivation of the change in refection 
or transmission probability. Use Eqs.~(\ref{eq:R}), (\ref{eq:del-p}), (\ref{eq:delR-2}),
and (\ref{eq:dp}) to write 
\begin{equation}
\delta R \approx  -4 \xi \eta^2\,e^2\, v_0^2\, R\,.
\label{eq:delR-3}
\end{equation}
Given that $\xi$ and $\eta$ are constants of order one, this estimate approximately agrees with
Eq.~(\ref{eq:delR}).

\section{Radiative Corrections to Quantum Tunneling}
\label{sec:tunnel}

Now we return to the situation illustrated in Fig.~\ref{fig:V}, but where the incident energy of the
particle is below the maximum of the potential barrier, $E_0 < V_0$. The transmission, or tunneling 
probability $T$ may again be found in nonrelativistic quantum mechanics by solving the
 Schr{\"o}dinger equation. In many cases, the result is accurately given by the WKB approximation,
 which leads to the result
 \begin{equation}
T \approx \exp\left(-\int_{x_1}^{x_2} \sqrt{2 m [V(x) -E_0]}\, dx \right) \,,
\label{eq:WKB}
\end{equation}
where $x_1$ and $x_2$ are the classical turning points at which $E_0= V(x)$. This is typically
a good approximation when the tunneling probability is small, $T \ll 1$.

The one loop radiative correction was given by Flambaum and  Zelevinsky~\cite{FZ99}, who
show that it results in an increase in tunneling probability. This increase is the same as would
arise if the potential were shifted by $V(x) \rightarrow V(x) +\delta V(x)$, where the potential
shift is
\begin{equation}
 \delta V(x)=  \nabla^2 V(x)\; \frac{e^2}{12\pi^2 m^2}\;\ln\left(\frac{m}{V_0}\right) \,,
 \label{eq:FV}
 \end{equation}
 where $V_0$ is the maximum value of $V(x)$.
For a potential such as that illustrated in Fig.~\ref{fig:V}, $\nabla^2 V(x) < 0$ near the maximum of the
potential, so $\delta V(x) < 0$, and the tunneling probability increases. This one loop effect
also arises from the vertex correction, Fig.~\ref{fig:vertex}, just as did the effect discussed in
Sec.~\ref{sec:rad}. Flambaum and  Zelevinsky introduce an infrared cutoff at a scale of $V_0$, the height of
the potential, so the logarithmic terms in Eqs.~(\ref{eq:MV}) and (\ref{eq:FV}) have the same origin.

If we approximate the form of the potential near its maximum as
\begin{equation}
V(x)  \approx -\frac{1}{2} V_0\, \left(\frac{x}{a}\right)^2+V_0\,,
\end{equation} 
then $ \nabla^2 V(x) \approx -V_0/a^2$. Now we have the estimate
\begin{equation}
\delta V(x) \approx - \beta\, \frac{e^2\, V_0}{m^2\, a^2}\, ,
\label{delta V}
\end{equation}
where $\beta = [\ln(m/V_0)]/(12\pi^2)$ is a positive constant which is expected to lie in the range 
between about $0.1$ and $0.01$, if $m$ is the electron mass. For example, if 
$1{\rm eV} < V_0 <  10^5 {\rm eV}$, then $0.014 < \beta < 0.11$.

Now we wish to give a heuristic derivation of Eq.~(\ref{delta V}) based upon the effects of vacuum
electric field fluctuations. However, there is a conceptual problem of defining the tunneling time
of a quantum particle. This issue has been much discussed in the literature. See 
Refs.~\cite{HS89,LM94} for review articles with extensive lists of references. The origin of
the ambiguity lies in the fact that localized quantum particles are described by wave packets
which can change shape as they pass under a potential barrier. A related ambiguity arises
for electromagnetic wave packets in a dispersive material. For our purposes, an order of magnitude
estimate for the tunneling time will be sufficient. If $V_0$, $E_0$ and $V_0-E_0$ are all of the same
order of magnitude, then one expects that this time will not be dramatically different from the time
required for a free particle of energy  $E_0 = \frac{1}{2} m v_0^2$ to travel a distance $a$, which
is $a/v_0$. That is, we expect $\tau \approx a/v_{\rm eff}$, where the effective speed $v_{\rm eff}$
is of order $v_0$. This expectation is supported by several explicit proposals, including one by 
B{\"u}ttiker and Landauer~\cite{BL88}, who suggest 
\begin{equation}
v_{\rm eff} \approx \sqrt{\frac{2(V_0-E_0)}{m}}\,,
\end{equation}
and a proposal by Davies~\cite{Davies04}
\begin{equation}
v_{\rm eff} \approx \frac{2 v_0}{2+V_0/E_0}\,.
\end{equation}

We will use an argument similar to that in Sec.~\ref{sec:eflucts}, where the electron is subjected
to an electric field of magnitude $\bar{\cal{E}} =\eta/\tau^2$, and a force of order $e \bar{\cal{E}}$.
This force does work of order $e\, a\, \bar{\cal{E}}$, and causes a momentum change of order
$e \bar{\cal{E}} \, \tau$, whose signs may be either positive or negative.
Here we will use the fact that if the force is in the direction of motion of the electron, it slightly decreases
the transit time and hence increases $\bar{\cal{E}}$ and the work. A force in the backwards direction
has the opposite effect. Let $v_+$ and $\tau_+$ be the effective speed and transit time when the force
is in the forward direction, and $v_-$ and $\tau_-$ be the corresponding quantities for a backward
force fluctuation. Then
\begin{equation}
v_+=v_{\text{eff}}+\frac{e\, \bar{\cal{E}}\, \tau_+}{2m}=v_{\text{eff}}+\frac{e\eta}{2m\tau_+}
\end{equation}
and
\begin{equation}
 \tau_+=\frac{a}{v_+}=\frac{a-\eta\frac{e}{2m}}{v_{\text{eff}}}=
 \frac{a}{v_{\text{eff}}} \left(1-\eta\frac{e}{2ma}\right)\,.
\end{equation}
(The factor of $1/2$ in the change in $v_+$ comes from taking an average velocity when the change in
momentum is given  by Eq.~(\ref{eq:del-p}).)
Similarly,
\begin{equation}
 \tau_-=\frac{a+\eta\frac{e}{2m}}{v_{\text{eff}}}=\frac{a}{v_{\text{eff}}}
 \left(1+\eta\frac{e}{2ma}\right) \,.
\end{equation}
The average change in kinetic energy due to a forward force fluctuations is
\begin{equation}
 \delta E_+ =\frac{ea\eta}{\tau_+^2}=\frac{ea\eta ~v_{\text{eff}}^2}{a^2}
 \left(1+\eta\frac{e}{ma}\right)\,,
\end{equation}
and that for the backward direction is
\begin{equation}
 \delta E_-=-\frac{ea\eta}{\tau_-^2}=-\frac{ea\eta ~v_{\text{eff}}^2}{a^2}
 \left(1-\eta\frac{e}{ma}\right)\,.
\end{equation}
The averaged change in energy is then
\begin{equation}
\delta E=\delta E_++\delta E_-=2\eta^2\frac{eav_{\text{eff}}^2}{a^2}\frac{e}{ma}=2\eta^2\frac{e^2v_{\text{eff}}^2}{a^2m}\sim
   \frac{e^2}{m^2}\frac{E_0}{a^2}\,.
\end{equation}
The net effect is an average increase in electron energy during the tunneling process, which is 
approximately equal in magnitude to the decrease in potential given by Eq.~(\ref{delta V}).
Within the WKB approximation, Eq.~(\ref{eq:WKB}) reveals that both correspond to the same
increase in tunneling probability. Thus our heuristic derivation based upon vacuum electric
field fluctuations agrees with the result of Flambaum and  Zelevinsky~\cite{FZ99}.

\section{Nonperturbative Effects of Electric Field Fluctuations}
\label{sec:nonpert}

In the previous sections, we have discussed one loop corrections to potential scattering, that is,
effects which may be calculated from perturbation theory. These effects are small compared to the
tree level transmission rates, and in our heuristic treatment, are estimated from the mean square
of the time averaged electric field, given by Eq.~(\ref{eq:E2}). Now we wish to turn to a discussion
of the effects of large electric field fluctuations. The probability of such fluctuations is determined
by Eq.~(\ref{eq:pdf}). Consider the situation discussed in Sec.~\ref{sec:tunnel}, where an electron
has an incident energy less than the height of the potential barrier. A sufficiently large electric
field fluctuation could temporarily give the electron enough energy to fly over the barrier. This 
extra energy is likely to be taken away by a fluctuation of the opposite sign, but this does not matter
if the particle is now on the far side of the barrier. Assume a temporary increase in electron energy 
of order
\begin{equation}
\delta E = e\, \bar{\cal E}\, a = e\,  a\, \tau^{-2}\, \chi \,,
\end{equation}
where $\delta E > V_0 - E_0$. We estimate that the electron flies over the barrier at a speed of order 
\begin{equation}
v_{\rm eff} \approx \sqrt{\frac{2( \delta E - V_0 +E_0)}{m}}
\end{equation}
and in a time of order $\tau \approx a/v_{\rm eff}$. We may combine these expressions and solve
for $\chi$ to find
\begin{equation}
\chi = \frac{a\,m}{2 e} + \frac{\tau^2}{e\, a} \,(V_0-E_0) > \chi_0  \,,
\label{eq:chi}
\end{equation}
where $\chi_0 = {a\,m}/(2 e) \,.$
Note that $\chi_0$ is somewhat larger than the width of the barrier $a$ measured as a multiple of 
the electron Compton wavelength, so we expect $\chi_0 \gg 1$. The lower bound on $\chi$ is never
actually attained because $\tau > 0$. However, if $V_0 - E_0 \ll m$, it may be possible to have
$\tau^2 (V_0 - E_0)/a \ll a m/2$, and hence $\chi$ nearly equal to $\chi_0$. We will assume this in our 
probability estimate.

The probability of a fluctuation in which $\chi \geq \chi_0$ is given by an integral of the probability
distribution function in Eq.~(\ref{eq:pdf})
\begin{equation}
{\cal P}(\chi \geq \chi_0) = \int_{\chi_0}^\infty P(\chi)\, d\chi \,. 
\label{eq:prob-int}
\end{equation} 
This integral may be evaluated in terms of the error function, and in the limit of large $\chi_0$, the
result is
\begin{equation}
{\cal P}(\chi \geq \chi_0) \approx \frac{\eta}{\chi_0 \sqrt{2 \pi}} \, \exp\left(-\frac{\chi_0^2}{2 \eta^2} \right)
 =   \frac{2 \, e\,\eta}{ \sqrt{2 \pi}\,a \,m} \, \exp\left(-\frac{a^2\, m^2}{8 \eta^2\, e^2} \right) \,. 
 \label{eq:nonpert}
\end{equation} 
We can view this result as a nonperturbative contribution to the transmission probability $T$.
Its nonperturbative character is demonstrated by the appearance of $e^2$ in the denominator
of an exponential. 

As noted above, $a\, m$ is the barrier width as a multiple of the electron Compton 
wavelength, so we expect $a\, m \gg1$ and hence the exponential factor in Eq.~(\ref{eq:nonpert})
to be very small. Another feature of this result which should be noted is the lack of explicit
dependence upon the barrier height, $V_0$.  This arises from our approximation in writing
Eq.~(\ref{eq:prob-int}) as an integral with a lower bound of $\chi_0$, which is independent of
$V_0$. This approximation is at best valid when $V_0 \ll m$, so a nonrelativistic
particle can fly over the barrier, and will fail for larger values of $V_0$. Within this approximation,
this  nonperturbative contribution is independent of the barrier height, although it depends strongly 
upon the barrier width.

\section{Summary and Discussion}
\label{sec:sum}

In this paper, we have reviewed some previous results on one loop QED corrections to quantum
potential scattering in one space dimension, including both above barrier scattering and quantum
tunneling. In both cases, the one loop correction increases the transmission probability. We have 
argued that the order of magnitude of this increase can be obtain from very simple arguments based
upon vacuum fluctuations of the quantized electric field. The basic idea is that there is a characteristic
transit time $\tau$ for the electron to pass through the potential, and this time defines a characteristic
magnitude for an electric field fluctuation, given by Eq.~(\ref{eq:E2}). This field fluctuation leads to
a temporary change in the energy and momentum of the electron, Although the sign of this change
can be either positive or negative, the average effect is an increase in transmission probability.

This argument gives a concrete physical meaning to time averaged quantum fields, such as defined
in Eq.~(\ref{eq:t-ave}). Both the width and shape of the potential determine the form of the test
function $f_\tau(t)$. A different model in the context of nonlinear optics was presented in 
Ref.~\cite{BDFS14} and led to a similar interpretation of the test function. 

In Sec.~\ref{sec:nonpert}, we used the probability distribution for electric field fluctuations, Eq.~(\ref{eq:pdf}),
to propose a nonperturbative effect of these fluctuations on quantum tunneling. Although this effect is
very small, it does hint that situations with different probability distributions could produce larger effects.
One such situation might be the effects of quantum stress tensor fluctuations, whose probability distributions
were obtained in special cases in Refs.~\cite{FFR10,FFR12}, and can decrease more slowly than a Gaussian.
The possible application of this effect to quantum tunneling is currently being investigated.

 \begin{acknowledgments}
This work was supported in part by the National Science Foundation under Grant PHY-1205764.
\end{acknowledgments}

\end{document}